\documentclass[
  aps,
  prb,
  twoside,
  twocolumn,
  floatfix
]{revtex4}
\usepackage{amsmath}
\usepackage{amssymb}
\usepackage{graphicx}
\usepackage{tabularx}
\usepackage{dcolumn}
\usepackage{longtable}
\usepackage{footnote}
\usepackage{appendix}

\begin{document}

\newcommand{\bec}{\begin{center}}
\newcommand{\ec}{\end{center}}
\newcommand{\be}{\begin{equation}}
\newcommand{\ee}{\end{equation}}
\newcommand{\beqn}{\begin{eqnarray}}
\newcommand{\eeqn}{\end{eqnarray}}
\newcommand{\bet}{\begin{table}}
\newcommand{\ent}{\end{table}}
\newcommand{\bib}{\bibitem}

%\baselineskip 4.2mm 

%\wideabs{

\title{
Anomalous nanoscale diffusion in Pt/Ti:
superdiffusive intermixing 
%simulated ion-sputtering in Pt/Ti
%Interface-anisotropy induced anomalous intermixing in Pt/Ti
%Interface-anisotropy induced asymmetry of intermixing in bilayers 
}

\author{P. S\"ule, M. Menyh\'ard} 
  \address{Research Institute for Technical Physics and Material Science,
www.mfa.kfki.hu/$\sim$sule, sule@mfa.kfki.hu\\
Konkoly Thege u. 29-33, Budapest, Hungary\\
%$^{\star}$ National Institute of Standards \& Technology, Gaithersburg, Maryland 20899\\
}
%\email{sule@mfa.kfki.hu}

%\date{\today}
\date{2007.10.24}

\begin{abstract}
Probing the anomalous nanoscale intermixing using molecular dynamics (MD) simulations
in Pt/Ti bilayer we characterize the superdiffusive nature of
interfacial atomic transport. 
In particular,
the low-energy ($0.5$ keV) ion-sputtering induced transient enhanced intermixing has been studied by MD simulations.
{\em Ab initio} density functional calculations have been used to check and reparametrize the employed heteronuclear interatomic potential.
We find a robust intermixing in Pt/Ti driven by nanoscale mass-anisotropy . The sum of the square of atomic displacements
$\langle R^2 \rangle$
asymptotically scales nonlinearly ($\langle R^2 \rangle \propto t^2$), where $t$ is the time of ion-sputtering, respectively
which is the fingerprint of superdiffusive features.
This anomalous behavior explains the high diffusity tail in the concentration profile obtained
by 
Auger electron spectroscopy depth profiling (AES-DP) analysis in Pt/Ti bilayer (reported in ref.: P. S\"ule, {\em et al.},
J. Appl. Phys., {\bf 101}, 043502 (2007)).
In  Ti/Pt bilayer a linear time scaling of $\langle R^2 \rangle \propto t$ has been found at the Ti/Pt interface indicating
the suppression of superdiffusive features.
%Hence we find that the succession of the film and substrate determines the
%asymptotics of the time scaling of interfacial broadening during ion-beam mixing
%in mass-anisotropic bilayers in accordance with the experimental findings.
These findings are inconsistent with the standard ion-mixing models.
Instead a simple accelerative effect of the downward fluxes of energetic
particles on the unidirectional fluxes of preferential intermixing of Pt atoms seems to explain the
enhancement of interfacial broadening in Pt/Ti.
Contrary to this in Ti/Pt the fluxes of recoils are in counterflow with intermixing Pt atoms and hence
slows down the nanoscale mass-effect driven ballistic preferential mobility of Pt atoms.

{\em PACS numbers:} 66.30.Jt, 61.80.Jh, 68.35.-p, 68.35.Fx, 66.30.-h, 68.55.-a \\
%{\scriptsize {\em Keywords:} intermixing, interdiffusion, anomalous atomic transport, interface, impurity diffusion, Auger depth profiling, sputtering, computer simulations, multilayer, ion-solid interaction, molecular dynamics, Ti/Pt, accelerative effects
%}
\end{abstract}
%}

\maketitle

\section{Introduction}

 The nanoscale production of nano-devices constitute a topic of high current interest
due to numerous potential applications \cite{Schukin,nano}.
The construction of sharp interfaces in the nanoscale, however,
faces many difficulties which requires the fundamental understanding
of nanoscale interfacial diffusion \cite{Philibert}.

 There are a growing number of evidences emerged 
in the last few years
that anomalous nanoscale broadening of interfaces or high diffusity tail in the impurity concentration profile
occur during ion-irradiation \cite{Abrasonis,Parascandola,Weber,Cardenas,Nieven,Sule_JAP07,CrSi}, sputter deposition of solids \cite{Buchanan} or during 
thin film reactions and reactive front propagation \cite{Vovk,Brockmann}. 
The anomalous nanoscale bulk diffusional effects
could be due to still not clearly established accelerative effects leading to anomalously fast and possibly athermal diffusion at interfaces or during impurity diffusion
\cite{Abrasonis,Parascandola,Weber,Cardenas,Nieven,Sule_JAP07,CrSi,Buchanan,Vovk,Bellon,Barna,Csik,CuTa,Delugas,Avasthi,Melis,Solmi,Sparks,Venezia,Slijkerman}.

  Anomalous atomic transport (AAT) in the bulk or on surfaces can be categorized
into few groups of phenomena:
 (i) Post-irradiation induced enhancement of impurity \cite{Abrasonis,Parascandola}
or dopant diffusion \cite{Cardenas,Delugas,Avasthi,Melis,Solmi,Sparks,Venezia,Slijkerman}
and transient enhanced interfacial broadening \cite{Weber,Nieven,Sule_JAP07,CrSi}.
Computer simulations reveal AAT during low-energy cluster and sigle-atomic deposition
\cite{ptonal}.
(ii)
The amplification of intermixing during thin film growth (sputter deposition) \cite{Buchanan,ptonal} or
forced alloying \cite{Bellon}.
(iii)
Anomalous growth rates and front or interface propagation \cite{Vovk,Brockmann}.
(iv) Superdiffusion, random walk, Levy flight \cite{anomalous,Levy,Luedtke,Michely,Sule_SUCI} or quantum tunneling diffusion \cite{Philibert,Bulou2} on solid surfaces.
(v) Ultrafast atomic mobility with the coherent or collective movement
of an ensemble of atoms (not cluster diffusion) \cite{ptonal,Nordlund_Nature,pbsi,Bulou}.

  The physics behind these processes is not clearly established yet hence
theoretical works are needed to model AAT which could explain
the relation between the diversity of such processes.
This is because AAT could not simply be explained by the conventional theories
of atom movements, such as vacancy, interstitial diffusion, 
simple atomic site exchanges or hopping mechanisms \cite{Philibert,Michely}.
In the present paper we show that a simple explanation, such as atomic
mass anisotropy induced amplification of ion-intermixing explains the occurrence of AAT in
the prototypical mass-anisotropic system of Pt/Ti.

 Computer atomistic simulations also reveal the occurrence of
the enhancement of atomic transport upon ion-bombardment of various
solids beyond a level explained by radiation enhanced diffusion \cite{Sule_JAP07,Nordlund_Nature,Sule_PRB05}.
The characterization and understanding of the nanoscale amplification or weakening
of
atomic transport in the bulk, at interfaces and on the surface could be an important ingredient of
the efficient production of nanostructures \cite{Vovk,Erdelyi}.
In particular, it has been shown that nanoscale mass anisotropy 
influences seriously the sharpness of anisotropic interfaces \cite{Sule_JAP07,Sule_PRB05}
as well as surface morphology \cite{Sule_SUCI,Sule_NIMB04_2} and adatom yield \cite{Sule_SUCI}.

 Under forced conditions (such as low-energy ion-sputtering or ion-beam
deposition, ball milling)
otherwise not easily observable anomalous atomic transport processes can be amplified
and could be detected.
Such conditions have widely been applied in the last decades to
force e.g. intermixing or alloying between immiscible elements \cite{Bellon,CuTa,Nastasi,AverbackRubia,Menyhard,Vazques}.
The athermal broadening of the interface in strongly intermixing ion-irradiated bilayers such as Ni/Al and Cr/Al have been found which are inconsistent with the 
standard ion beam mixing models, such as ballistic, thermal spike or
radiation enhanced diffusion \cite{Weber}.
In these materials, such as Ni/Al or Co/Al, the exotermic solid state reaction can even
lead to extremely fast burn rates \cite{Vovk,shock}.
The anomalously strong asymmetric ion-beam mixing has been found recently in Cr/Si multilayer
using focused ion beam \cite{CrSi}.

 Moreover, transient enhanced diffusion (TED) and intermixing have also been reported in post-annealed
dopant implanted semiconductors \cite{Delugas,Melis,Solmi,Sparks,Venezia,Slijkerman}.
TED occurs when the depth distribution of the dopant exceeds the ion range \cite{Abrasonis}.
Using
low-energy ion-sputtering \cite{Nastasi,Gnaser,IBAD} and Auger electron depth profiling analysis \cite{Menyhard} it has
been concluded recently  
that transient enhanced intermixing
could occur in the film/substrate bilayer Pt/Ti while no such behavior is seen in
the Ti/Pt bilayer, hence the magnitude of intermixing might depend
on the succession of the film and the substrate
\cite{Sule_JAP07}.

 Anomalously long interdiffusion depths have been found in various diffusion
couples \cite{Abrasonis,Weber,Buchanan,Delugas,Melis,Solmi,Sparks,Venezia,Slijkerman}.
Transient enhanced intermixing has been reported in nonstochiometric AlAs/GaAs quantum well
structures \cite{GaAs} or in AsSb/GaSb superlattices \cite{AsSb} and
has been attributed to vacancy or self-interstitial supersaturation in
annealed samples.
It has also been reported that in few cases
anomalous intermixing is neither driven by bulk diffusion parameters nor by thermodynamic forces (such as heats of alloying) \cite{Buchanan} or nor by heats of mixing \cite{Sule_PRB05,Sule_NIMB04}.
It has also been found that during low-energy ion-bombardment of bilayers
the intermixing length (and the mixing efficiency) scales nonlinearly with the mass anisotropy
(mass ratio)
leading to the abrupt increase of the mixing efficiency in mass-anisotropic bilayers
\cite{Sule_PRB05}.

 In the present work, computer atomistic simulations have been carried out
to explain the occurrence of the ion-sputtering induced high diffusity tail in the concentration profile of the film/substrate system of Pt/Ti.
   The enhanced intermixing in Pt/Ti reported in ref. \cite{Sule_JAP07} is attempted to interpret as a superdiffusive atomic transport
process 
since it fulfills the most important condition of superdiffusion:
the square of atomic displacements scales nonlinearly with the time of IM \cite{anomalous}.
Superdiffusion has only been reported until now on solid surfaces
\cite{Levy,Luedtke,Michely} and no reports have been found
for bulk superdiffusion except for ultralight particles such as the migration of H in metals \cite{Philibert}.
%The most recently reactive diffusion dynamics has been interpreted
%as a superdiffusive process during the front propagation of interfaces \cite{Brockmann}.

  We find the long range atomic transport of Pt in Ti is highly unusual and which could be explained as
a mass anisotropy driven superdiffusive intermixing (SIM) process.
Moreover, we conclude that SIM occurs due to the accelerative effect of
unidirectional fluxes of energetic particles.

\section{The setup of the atomistic simulations}

 Classical molecular dynamics simulations have been used to simulate the ion-solid interaction
(using the PARCAS code \cite{Nordlund_ref}).
%Here we only shortly summarize the most important aspects.
A variable timestep
and the Berendsen temperature control is used to maintain the thermal equilibrium of the entire
system. \cite{Allen}. 
The global coupling to the heat bath can be adjusted by the so called
Berendsen temperature which we set to $70$ K.
Temperature controll has been applied at the cell borders of the simulation cell
to maintain constant temperature conditions.
The bottom layers
are held fixed in order to avoid the rotation of the cell.
Since the z direction is open, rotation could start around the z axis.
The bottom layer fixation is also required to prevent
the translation of the cell.
Periodic boundary conditions are imposed laterarily and a free surface is left for the ion-impacts.
The simulation uses the Gear's predictor-corrector algorithm to calculate
atomic trajectories \cite{Allen}.
%The temperature of the atoms in the
%outermost layers was softly scaled towards the desired temperature to provide temperature control and ensure
%that the pressure waves emanating from cascades were damped at the borders.
%The lateral sides of the cell are used as heat sink (heat bath) to maintain the thermal equilibrium of the entire
%system \cite{Allen}.
The detailed description of other technical aspects of the MD simulations are given in refs. \cite{Nordlund_ref} and \cite{Allen} and details specific to the current system in recent
communications \cite{Sule_JAP07,Sule_PRB05,Sule_NIMB04}.

 Our primary purpose is to simulate the conditions occur during ion-sputtering \cite{Sule_JAP07,Gnaser}
and Auger electron spectroscopy depth profiling analysis (AES-DP) \cite{Sule_JAP07}
using molecular dynamics simulations \cite{Allen}.
Recently, MD simulation has been used to simulate ion-sputtering induced surface roughening
\cite{Sule_NIMB04_2,Karolewski,Thijsse}.
Following our previous work \cite{Sule_JAP07}
we ion bombard the bilayers Pt/Ti and Ti/Pt 
with 0.5 keV Ar$^+$ ions repeatedly (consecutively) with a time interval of 10-20 ps between each of
the ion-impacts at 300 K
which we find
sufficiently long time for most of the structural relaxations and the termination of atomic mixing, such
as sputtering induced intermixing (IM) \cite{Sule_NIMB04}.
Since we focus on the occurrence of transient intermixing atomic transport processes,
the relaxation time of $10-20$ ps should be appropriate for getting adequate information
on transient enhanced intermixing.
Pair potentials have been used
 for the interaction of the Ar$^+$ ions with the metal atoms derived using
{\em ab initio} density functional calculations.

 The initial velocity direction of the
impacting ion was $10$ degrees with respect to the surface of the crystal (grazing angle of incidence)
to avoid channeling directions and to simulate the conditions applied during ion-sputtering \cite{Sule_JAP07}. 
The impact positions have been randomly varied on the surface of the film/substrate system and the azimuth angle $\phi$ (the direction of the ion-beam).
In order to simulate ion-sputtering a large number of ion irradiation are
employed using script governed simulations conducted subsequently together with analyzing
the history files (movie files) in each irradiation steps.
In this article results are shown up to $150$ ion irradiation (in a similar way to that given in ref. \cite{Sule_JAP07}).
The impact positions of the $100-150$ ions are randomly distributed
over a $20 \times 20$ \hbox{\AA}$^2$ area on the surface.

 The volume of the cubic simulation cell is $110 \times 110 \times 90$ $\hbox{\AA}^3$ including
$\sim 57000$ atoms (with 9 monolayers (ML) film/substrate).
The film and the substrate are $\sim 20$ and $\sim 68$ $\hbox{\AA}$ thick, respectively.
 The setup of the simulation cell, in particular the $20$ $\hbox{\AA}$ film thickness is assumed to be appropriate for simulating broadening. Our experience shows that the variation of the film thickness does not affect the final result significantly, except if ultrathin film is used (e.g. if less than
$\sim 10$ $\hbox{\AA}$ thick film). At around $5$ or less ML thick film surface roughening could affect
mixing \cite{Sule_NIMB04_2}.

  The (111) interface of the fcc crystal is parallel to (0001) of the hcp
crystal
and the close packed directions are parallel.
The interfacial system
has been created as follows:
the hcp Ti is put by hand on the (111) Pt bulk (and vice versa) and various structures with different lateral sizes have been probed
and are put together with fixed orientation mentioned above. 
This must be done to avoid built in stress in the initial stucture.
Therefore we put together slabs of the film and the substrate with
different width while keeping (111) interfacial orientation. This is because the lattice mismatch
is sensitive to the
relative positions of the atoms at the interface. In order to minimize
lattice misfit in the initial structure the relative lateral positions of the
film and the substrate has been varied and the interfacial system with the
smallest misfit strain has been selected. In fact this has been monitored
via the starting average temperature of the system which indicates us how
the generated strucure is relaxed. The closer this temperature to 0 K the more
relaxed the system is.
The remaining misfit is properly minimized below $\sim 6 \%$ during the relaxation
process so that the Ti and Pt layers keep their original crystal structure and we
 get an
atomically sharp interface.
During the relaxation (equilibration) process the temperature is softly scaled down
to zero and a sufficiently relaxed structure has been obtained.
According to our practice we find that during the temperature scaling down the 
structure
becomes sufficiently relaxed.
Then the careful heating up 
of the system to $300$ K has also been carried out.
We believe that the lattice mismatch is minimized to the lowest possible level
and we are convinced that no serious built-in stress remained in the
simulation cell which could cause the explored anomalous atomic transport behaviors.

\begin{table}[t]
\caption[]
{
The parameters used in the tight binding potential given in Eqs. (1)-(2) \cite{CR}
}
\begin{ruledtabular}
\begin{tabular}{lccccc}
%\hline
 & $\xi$ & q & A & p & $r_0$  \\
\hline
 Ti  & 1.416  & 1.643  & 0.074 & 11.418 & 2.95 \\
 Pt  & 2.695 & 4.004  & 0.298 & 10.612 & 2.78   \\
 Ti-Pt & 4.2 & 2.822 & 0.149  & 11.015 & 2.87   \\
%\hline
%---------------------------------------------------------------
\end{tabular}
\end{ruledtabular}
\footnotetext[1]{
The parameters of the crosspotential have been obtained
as follows \cite{ZBL}:
For the preexponentials $\xi$ and $A$ we used the
harmonic mean $A_{TiPt}=(A_{Ti} \times A_{Pt})^{1/2}$
($\xi$ has been fitted to the {\em ab initio} curve shown
in Fig. \ref{potential}),
for $q$ and $p$ we use the geometrical averages:
$q_{TiPt}=(q_{Ti} + q_{Pt})/2$. The first neighbor distance of the
Ti-Pt potential is given also as a geometrical mean of $r_0=(r_0^{Pt}+r_0^{Ti})/2$.
}
\label{table}
\end{table}

 We used a tight-binding many body potential
on the basis of the second moment approximation (TB-SMA) to the density of states \cite{CR}, to describe interatomic interactions.
 Within the TB-SMA, the band energy
the attractive part of the potential reads,

%------------------------------------------------------
\begin{figure}[hbtp]
\begin{center}
\includegraphics*[height=5cm,width=6cm,angle=0.]{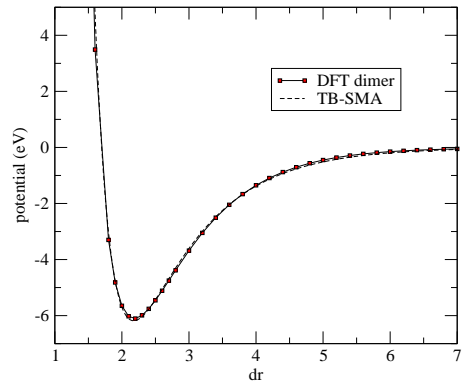}
\caption[]{
The crosspotential energy (eV) for the Ti-Pt dimer as a function of
the interatomic distance ($\hbox{\AA}$) obtained by
the first-principles PBE/DFT method. For comparison the fitted interpolated TB-SMA potential
is also shown calculated for the Ti-Pt dimer.
}
\label{potential}
\end{center}
\end{figure}
%------------------------------------------------------

\be
 E_b^i=-\biggm[ \sum_{j, r_{ij} < r_c} \xi^2 exp \biggm[-2q \biggm(\frac{r_{ij}}{r_0}
-1 \biggm) \biggm] \biggm]^{1/2},
\ee
where $r_c$ is the cutoff radius of the interaction and $r_0$ is the first neighbor distance (atomic size parameter).
The repulsive term is a Born-Mayer type phenomenological core-repulsion term:
\be
E_r^i=A \sum_{j, r_{ij}<r_c} exp \biggm[-p \biggm(\frac{r_{ij}}{r_0}-1 \biggm) \biggm].
\ee
The parameters ($\xi, q, A, p, r_0$) are fitted to experimental values of the cohesive energy,
the lattice parameter, the bulk modulus and the elastic constants $c_{11}$, $c_{12}$ and $c_{44}$ \cite{CR} and which are given in Table 1.
$r_{ij}$ is the internuclear separation between atoms $i$ and $j$.
The total cohesive energy of the system is
\be
E_c=
\sum_i^{Nat} (E_r^i+E_b^i),
\ee
where $Nat$ is the number of atoms in the system.

%This tight-binding approach is formally analogous to the embedded-atom method (EAM) \cite{EAM}.
The TB-SMA potential gives a good description of lattice vacancies, including atomic migration
properties and a reasonable description of solid surfaces and melting \cite{CR}.
In ref. \cite{Sule_JAP07}.
it has been shown that the TB-SMA potential gives the reasonable description of IM in Pt/Ti
and gives interfacial broadening comparable with AES-DP measurements.
%Since the present work is mostly associated with the elastic properties,
%melting behaviors, interface and migration energies, we believe the model used should be suitable for this study.
Cutoff is imposed out of the 2nd nearest neighbors when the
interatomic interactions have been calculated which we find sufficient for
simulating ion-mixing \cite{Sule_PRB05,Sule_NIMB04}.
Simulations have also been conducted using larger cutoff distances (up to 4th neighbors),
however,
no serious change has been observed in the final results.
%This amounts to the maximum interatomic distance of $\sim 6$ \hbox{\AA}.

 An interpolation scheme has been employed 
 for the crosspotential Ti-Pt \cite{Sule_JAP07,Sule_PRB05,Sule_SUCI,ZBL}.
The employed potentials and the interpolation scheme for heteronuclear interactions
have successfully been used for MD simulations \cite{Sule_PRB05,Stepanyuk2,Goyhenex,Levanov}.
The Ti-Pt interatomic crosspotential of the TB-SMA potentials \cite{CR} type has been fitted recently to the experimental
 heat of
mixing of the corresponding alloy system \cite{Sule_NIMB04,Sule_NIMB04_2}.
The scaling factor $r_0$ (the heteronuclear first neighbor distance) is given as the average of the elemental first neighbor distances.

 In this paper we use instead of our recent fit of the Ti-Pt potential \cite{Sule_NIMB04}
a more sophisticated potential.
 The crosspotential energy has been calculated for the Ti-Pt dimer
  using {\em ab initio} local spin density functional calculations \cite{G03} together with quadratic convergence self-consistent field (SCF) method.
The G03 code is well suited for molecular calculations, hence
it can be used for checking pair-potentials.
The interatomic potential $V(dr)$ between two atoms
is defined as the difference of total energy at an interatomic separation $dr$
and the total energy of the isolated atoms 
\be
 V(dr)=E(dr)-E(\infty).
\ee
The Kohn-Sham equations (based on density functional theory, DFT) \cite{KS} are solved in an atom centered Gaussian basis set and the core electrons
are described by effective core potentials
%We used effective core potentials
(using the LANL2DZ basis set) \cite{basis}
and
we used the Perwed-Burke-Ernzerhof (PBE) gradient corrected exchange-correlation potential \cite{PBE}.
Fist principles calculations based on
density functional theory (DFT) have been applied in various fields
in the last few years \cite{DFT_Sule}.

  The obtained profile is plotted in Fig. \ref{potential}
together with our interpolated TB-SMA potential for the Ti-Pt dimer.
We find that our interpolated TB-SMA potential
nearly perfectly matches the ab initio one hence we are convinced
that the TB-SMA model accurately describes the heteronuclear
interaction in the Ti-Pt dimer.
In fact we fitted only parameter $\xi=4.2$, which influences
the deepness of the potential well. The rest of the parameters 
are obtained using the interpolation scheme outlined in the
caption of Table 1.
We assume that this dimer potential is transferable for
those cases when the Pt atom is embedded in Ti.
This can be done because, as we outlined above, the interpolated
Ti-Pt potential properly reproduces the available experimental results
for the Ti-Pt alloy \cite{Sule_NIMB04}.

 The crossectional computer animations of simulated ion-sputtering can be seen in our web page \cite{web}.
Further details  are given in refs. \cite{Sule_JAP07,Sule_NIMB04,Sule_PRB05}.

\section{Results}

\begin{figure}[hbtp]
\begin{center}
\includegraphics*[height=4.3cm,width=5.5cm,angle=0.]{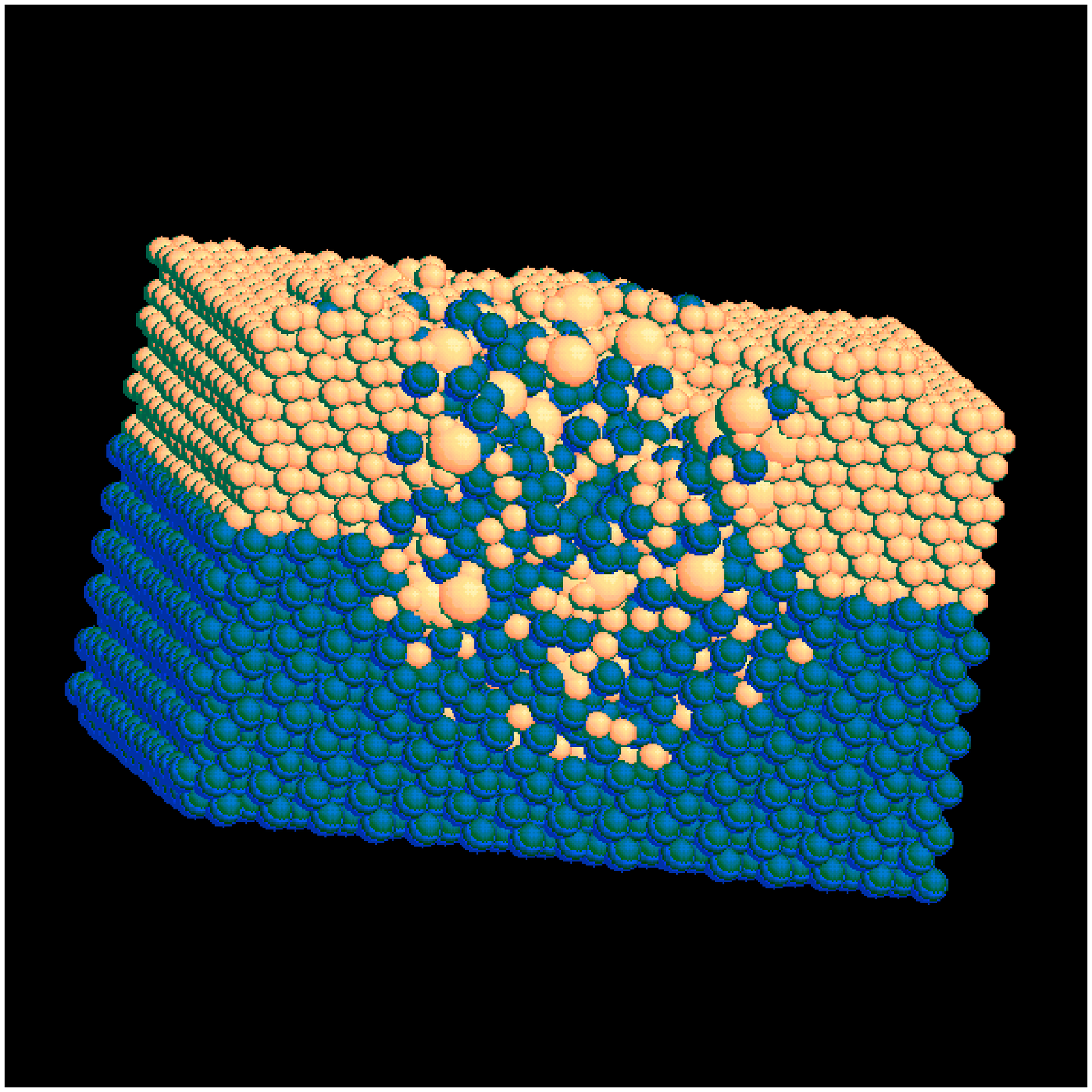}
\includegraphics*[height=4.3cm,width=5.5cm,angle=0.]{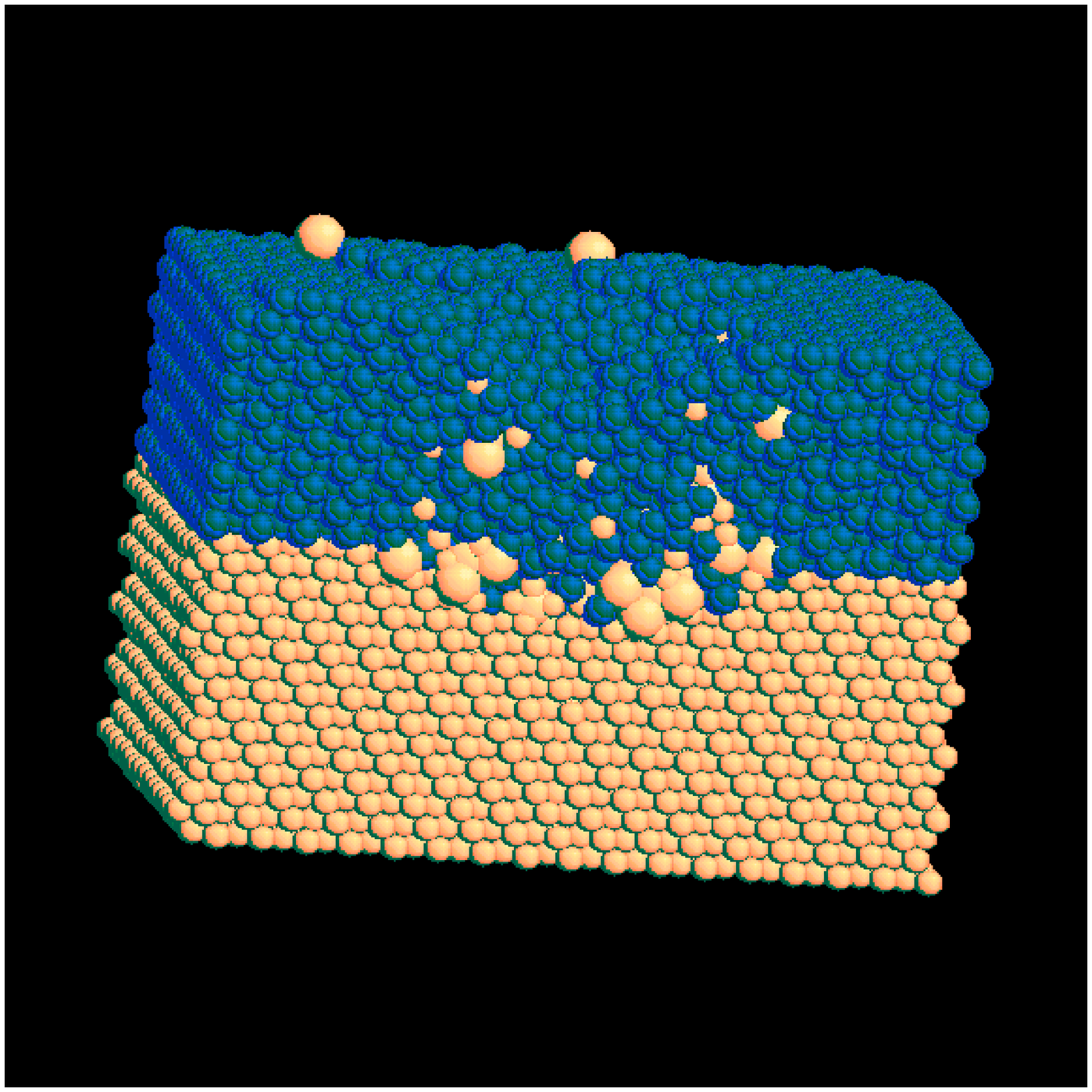}
\caption[]{
The cartoons of the simulation cells after $100$ ions bombardment
as a crossectional view (cut in the middle of the cell).
The incorporated Ar$^+$ ions are also shown as larger light spheres.
The smaller light and dark spheres denote the Pt and Ti atoms, respectively.
{\em Fig 2a}: Pt/Ti,
{\em Fig 2b}: Ti/Pt
}
\label{cartoon}
\end{center}
\end{figure}

  The cartoons of the simulation cells (crossectional slabs as a 3D view)
can be seen in Fig ~\ref{cartoon} which show the strong mixing at the interface
in Fig. 1a (Pt/Ti) and a much weaker mixing in Fig. 1b (Ti/Pt).

\begin{figure}[hbtp]
\begin{center}
\includegraphics*[height=6cm,width=7.5cm,angle=0.]{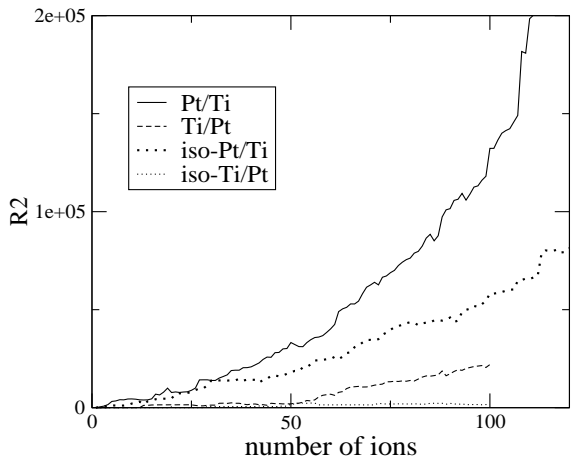}
\caption[]{
The simulated square of IM atomic displacements $\langle R^2 \rangle$ ($\hbox{\AA}^2$) in Pt/Ti, Ti/Pt and in Cu/Co as a function of the
ion-fluence (number of ions) obtained during the ion-sputtering of these bilayers at 500 eV ion energy (results are shown up to 100 ions).
The dotted lines (iso-Pt/Ti and iso-Ti/Pt) denote the results obtained for the artificial mass-isotropic
Pt/Ti and Ti/Pt bilayers, respectively.
%The ions are initiated from the surface.
}
\label{R2}
\end{center}
\end{figure}

  In Fig ~\ref{R2} the evolution of the sum of the square of atomic displacements (SD)   
\be
\langle R^2 \rangle= \sum_i^{N_{atom}} [{\bf r_i}(t)-{\bf r_i}(t=0)]^2,
\ee
of all intermixing atoms  
obtained by molecular dynamics simulations, where (${\bf r_i}(t)$ is the position vector of atom 'i' at time $t$, $N_{atom}$ is the total number of atoms included in the sum), can be followed as a function of the ion fluence.
%------------------------------------------------------
%------------------------------------------------------
Lateral components ($x,y$) are excluded from
$\langle R^2 \rangle$ and only contributions from IM atomic displacements perpendicular to the layers
are included ($z$ components). We follow during simulations the time evolution
of $\langle R^2 \rangle$ which reflects the atomic migration through the interface (no other
atomic transport processes are included).
Note, that we do not calculate the mean square of atomic displacements (MSD) which is an
averaged SD over the number of atoms included in the sum (MSD=$\langle R^2 \rangle/N_{atom}$).
MSD does not reflect the real physics when localized events take place, e.g. when only few dozens of atoms
are displaced and intermixed.
In such cases the divison by $N_{atom}$, when $N_{atom}$ is the total number of the atoms in the simulation cell
leads to the meaningless $\langle R^2 \rangle/N_{atom} \rightarrow 0$ result when $N_{atom} \rightarrow \infty$,
e.g. with the increasing number of atoms in the simulation cell.
Also, it is hard to give the number of "active" particles which
really take place in the transient atomic
transport processes.
Hence we prefer to use the more appropriate quantity SD.
In Fig. ~\ref{R2} we present
$\langle R^2 \rangle$ as a function of the number of ion impacts $N_i$ (ion-number fluence).
$\langle R^2 \rangle (N_i)$ corresponds to the final value of
$\langle R^2 \rangle$
obtained during the $N_i$th simulation. The final relaxed structure of the simulation of the
$(N_i-1)$th ion-bombardment is used as the input structure for the $N_i$th ion-irradiation.
The
asymmetry of
 mixing can clearly be seen when $\langle R^2 \rangle (N_i)$
and the depth profiles given in ref. \cite{Sule_JAP07} are compared in 
Ti/Pt and in Pt/Ti.
%------------------------------------------------------
%------------------------------------------------------
The computer animations of the simulations \cite{web} together with the plotted broadening values
at the interface in 
ref. \cite{Sule_JAP07} also reveal the stronger
IM in Pt/Ti.
 Moreover we find the strong divergence of $\langle R^2 \rangle$ from linear scaling
for Pt/Ti while linear scaling has been found for Ti/Pt.

 As it has already been shown in ref. \cite{Sule_JAP07}
AES-DP found
 a relatively weak IM in Ti/Pt (the interface broadening $\sigma \approx 20$ $\hbox{\AA}$) whil
e an unusually high
IM occurs in the Pt/Ti bilayer ($\sigma \approx 70$ $\hbox{\AA}$).
MD simulations provide $\sim 20$ $\hbox{\AA})$ and $\sim 40$ 
$\hbox{\AA})$ thick interface after $200$ ion impacts, respectively.

\subsection{The effect of mass anisotropy}

 In order to clarify the mechanism of intermixing and to understand how much
the interfacial anisotropy influences IM,
  simulations have been carried out with atomic mass ratio $\delta=m_{Pt}/m_{Ti}$, where $m_{Pt}$ and $m_{Ti}$ are the atomic masses, is artificially set to $\delta
 \approx 1$ (mass-isotropic). We find that $\langle
R^2 \rangle$ is below the corresponding curve of Pt/Ti (see
Fig ~\ref{R2}, iso-Pt/Ti, dotted line, see also the corresponding animation \cite{web}).
The $\langle R^2 \rangle$ scales nearly linearily as a function of the number of ions
(and with $t$) for iso-Pt/Ti.
Hence the asymptotics of $\langle R^2 \rangle$ is sensitive to the effect of $\delta$.
We reach the conclusion that the mass-effect is robust and the magnitude of IM is
weakened significantly.
Actually the system undergoes the transition in the asymptotics of $\langle R^2 \rangle
\propto t^2 \rightarrow \langle R^2 \rangle \propto t$.
This finding together with our AES measurements (with the long-range tail shown in ref. \cite{Sule_JAP07}) confirms our recent results reported for
various bilayers in which a strong correlation has been obtained between
the experimental and simulated mixing efficiencies and mass anisotropy in various metallic
bilayers
\cite{Sule_PRB05}.
In that article we found that below a certain threshold mass ratio value ($\delta \le 0.33$)
the rate of intermixing increases abruptly \cite{Sule_PRB05}.
On the basis of the results obtained in this paper this
surprising interdiffusive behavior of bilayers
could be explained by the anomalous nature of mixing which can be tuned by
the mass-anisotropy (mass ratio) of the systems.
The experimentally observed mixing asymmetry can also partly be explained by
the mass effect.
%However, as it can be seen from Fig. \ref{R2}, the curve of iso-Pt/Ti
%exhibits stronger IM than that of Ti/Pt. Hence the remaining part of
%the enhnaced IM in iso-Pt/Ti must be due to still an unknown effect.

  {\em The interchange of atomic masses:} 
 To further test mass-effect on IM, we carried out simulations
for the Pt/Ti system in which the atomic masses have been interchanged (Ti possesses the atomic mass of Pt and vice versa) setting in
an artificial mass ratio (the inverse of the normal one).
We find that this artificial setup of atomic masses results in the suppression
of IM in Pt/Ti.
Moerover, if we interchange the masses in Ti/Pt, we find strong IM and nonlinear scaling of $
\langle R^2 \rangle$, while
we find a much weaker one with natural atomic masses.
%We conclude from this that the sign of the mass ratio $\delta$ plays a huge role
%in the appearance of the enhanced IM.
The effect of mass anisotropy seems to be decisive in the occurrence of
transient enhanced IM in Pt/Ti and of the asymmetry of IM.

\section{Discussion}

\subsection{The nonlinear scaling of $\langle R^2 \rangle$}

  In recent papers we explained single-ion impact induced intermixing 
governed by the mass-anisotropy parameter (mass ratio) in these bilayers
\cite{Sule_PRB05,Sule_SUCI}.
%We also present $\langle R^2 \rangle$ results for Cu/Co bilayer in Fig 2.
%The details of the simulation in Cu/Co is given elsewhere \cite{CuCo}.
%Ti/Pt and Pt/Ti are mass-anisotropic, while Cu/Co is
%a mass-isotropic bilayer.
%Cu/Co can be used as a reference system, being a mass-isotropic bilayer, 
%hence a weak rate of IM is expected for.
  It has already been shown in refs. \cite{Sule_PRB05,Sule_NIMB04} that the backscattering of the light hyperthermal particles (BHP)
at the mass-anisotropic interface leads to the increase in the energy density of the displacement cascade.
We have found that the jumping rate of atoms through the interface is seriously affected by the
mass-anisotropy of the interface when energetic atoms (hyperthermal particles) are present and which leads to the
preferential IM of Pt to Ti \cite{Sule_PRB05,Sule_NIMB04}.

 Although we find in accordance with ref. \cite{Sule_PRB05} that thermal spike
occurs in both systems with the lifetime of few ps,
however, we rule out thermal spike effects on IM in Pt/Ti. 
In particular, the observed insensitivity of IM to the choice of the heat of mixing $\Delta H$ in Ti/Pt \cite{Sule_NIMB04}
is in contrast with the thermal spike model \cite{Sule_PRB05}.
The appearance of $\langle R^2 \rangle \propto t^2$ scaling requires
the presence of hyperthermal particles which are present only during
the collisional cascade.
These particles are also present in Ti/Pt, however in Pt/Ti we find the further
acceleration of the hot atoms due to unknown origin.

  We reach the conclusion that there must be an accelerative force field which
speeds up few of the Pt particles to ballistic transport.
The process must be active during the cascade.
This is reflected by the divergence of $\langle R^2 \rangle$ from linear scaling in Fig ~\ref{R2} for
Pt/Ti.
%------------------------------------------------------
\begin{figure}[hbtp]
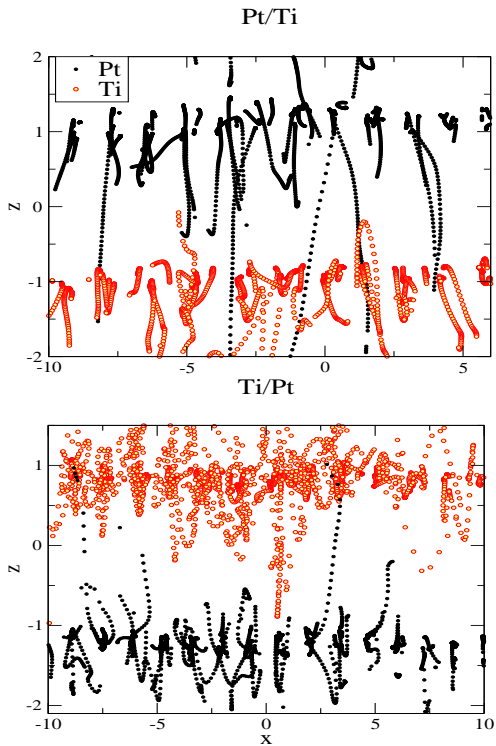

\begin{center}
\includegraphics*[height=4.9cm,width=6.5cm,angle=0.]{fig3a.eps}
\includegraphics*[height=4.9cm,width=6.5cm,angle=0.]{fig3b.eps}
\caption[]{
The crossectional view of a typical collisional displacement cascade at the interface
with atomic trajectories (two monolayers are shown at the interfaces as a crossectional
slab cut in the middle of the simulation cell)
in Pt/Ti (upper panel) and in Ti/Pt (lower panel).
The positions of the energetic particles are collected up to 500 fs during
a $500$ eV single ion-impact event.
The vertical axis corresponds to the depth position given in $\hbox{\AA}$.
The position $z=0$ is the depth position of the interface.
$x$ is the horizonthal position ($\hbox{\AA}$).
}
\label{traject}
\end{center}
\end{figure}
%------------------------------------------------------
Interdiffusion takes place via ballistic jumps (ballistic mixing), when $\langle R^2 \rangle$
 grows
asymptotically as $N^2$, where $N$ is the number fluence
(the same asymptotics holds as a function of ion-dose or ion-fluence).
This can clearly be seen in Fig. ~\ref{R2} for Pt/Ti.
The horizontal axis is proportional to the time of ion-sputtering, hence
$\langle R^2 \rangle
\propto t^2$ which is the time scaling of ballistic atomic transport
\cite{anomalous}.

In our particular case we follow the time evolution of the simulation cell after each of the ion impacts until
$t \sim 10-20$ ps which we find sufficient time for the evolution of $\langle R^2 \rangle$.
Anyhow, above this $t$ value the asymptotics of $\langle R^2 \rangle (t)$ is invariant
to the choice of the elapsed time/ion-bombardment induced evolution of $\langle R^2 \rangle (t)$.
Hence the transformation between ion-fluence and time scale is allowed.
$\langle R^2 \rangle \propto t^2$ and $\langle R^2 \rangle \propto t$
time scalings have been found even for the single-ion impacts averaged for few events (when $\langle R^2 \rangle (t)$ is plotted only for a single-ion impact) for Pt/Ti and Ti/Pt, respectively.
The $\langle R^2 \rangle \propto t^n$, scaling, where $n \ge 2$, used to be
considered as the signature of
anomalous diffusion (superdiffusion) in the literature \cite{anomalous}.
Such kind of time scaling has been repoted until now during
the random walk or flight of particles and clusters on solid surfaces
\cite{anomalous,Levy,Luedtke,Michely,Sule_SUCI}.
These processes are inherently athermal due to the vanisingly small activation energy of
surface diffusion.
We would like to show that it might also be the case that
transient IM takes place in Pt/Ti which resembles in many respect
the superdiffusive atomic transport processes known on solid surfaces \cite{anomalous}.

 In ref. \cite{Sule_JAP07} the concentration profiles
measured by
Auger electron spectroscopy (AES) depth profiling analysis
have been reported.
The obtained results are in agreement with the findings presented in this article.
However, in that paper it has not been realized that
the fingerprint of superdiffusive feature
of IM is detected by AES as a diffusity tail in the concentration profile of Pt
at the Pt/Ti interface in the Pt/Ti bilayer.
No such tail occurs in the concentration profile of Ti/Pt shown in ref. \cite{Sule_JAP07} where the profile
can be characterized by "normal" $erf$ functions.
Hence we find that the succession of the film and substrate could determine
the magnitude of IM (the asymmetry of IM).

  No such ballistic behavior can be seen for Ti/Pt in Fig ~\ref{R2}.
In Ti/Pt we find $\langle R^2 \rangle \propto t$ time scaling.
The mean free path of the energetic particles are much shorter in Ti/Pt. This can be seen
qualitatively in Fig ~\ref{traject} if we compare the length of the atomic trajectories for Pt
 between
upper and lower panels of Fig ~\ref{traject}.
In the plot of Pt/Ti in Fig ~\ref{traject} we can see ballistic trajectories
which result in the superdiffusive spread of Pt atoms.

 The trajectories of the reversed Ti recoils can be seen in upper Fig. ~\ref{traject} and
no intermixing Ti atomic positions can be found in the upper panel of Fig ~\ref{traject} (Ti/Pt).
Although, Fig. ~\ref{traject} has no any statistical meaning, however,
the atomic trajectories are plotted from a typical cascade event hence some useful information
can be obtained for the transport properties of energetic Pt atoms.
In the lower panel of Fig ~\ref{traject} we can see the ballistic trajectories of intermixing hyperthermal Pt atoms (Pt/Ti).
The reversed Ti particles at the interface
and the weaker IM of Pt atoms to the Ti phase
result in
the weaker IM in Ti/Pt than
in the Pt/Ti system.
%This is nicely reflected in the $\langle R^2 \rangle$ of Ti/Pt which remains in the range of Cu/Co.
Hence Fig. ~\ref{traject} depicts us at atomistic level what we see in the more statistical quantity 
$\langle R^2 \rangle$. In Ti/Pt we see much shorter inter-layer atomic trajectories
while in Pt/Ti ballistic trajectories of Pt atoms can be seen (moving through the interface).

\subsection{The ballistic model: deposited energy}

The ballistic model and the ballistic regime in the collisional cascade can also be ruled out
 as the source of the
asymmetry of IM.
We calculated the magnitude of the total deposited energy $F_D$ in Pt and in Ti, and
get the values of $127.6$ and $148.5$ eV for Pt/Ti and for Ti/Pt, respectively
at $0.5$ keV incident ion energy.
Although the $F_D$ is larger in Ti/Pt, the intermixing is weaker in
this material.
Using TRIDYN calculations one can estimate the magnitude of the
deposited energy at the interface \cite{Nastasi,TRYDIN,MM_TRYDIN}.
We calculate few eV/$\hbox{\AA}$ at the interface and again $F_D$ is smaller
in the case of Pt/Ti.
We conclude from this that not the larger $F_D$ and the smaller stopping power of
Pt causes the anomalous IM in Pt/Ti.

 The larger the $E_d$ smaller the number of displaced atoms $n_d$ at the interface
which leads to smaller $F_D$ following the Kinchin-Pease formula of
$F_D=2 n_d E_d$ \cite{AverbackRubia}.
Although we have no values for $n_d$ in various materials, however,
the corresponding $E_d$ values for Pt ($33$ eV) and Ti ($19$ eV) can be found in text books \cite{Nastasi}.
One can see that the $E_d$ of Pt is much larger than that of Ti, which again suggest
that simply if we rely only on the ballistic model we expect larger IM in Ti/Pt.
This is because such a big difference in $E_d$ values  
would strongly suggest that much larger IM should occur in Ti/Pt 
than in Pt/Ti. Also we can expect that the number of Frenkel pairs
is much larger in Ti than in Pt 
under the same iradiation conditions.
The Kinchin-Pease formula should give somewhat larger $F_D$ for Ti/Pt.
However, we get comparable values
as obtained by MD and SRIM simulations.
Than we can conclude that the difference in the deposited energy at the interface can not be the reason of IM amplification
in Pt/Ti.

 Also, at low-energy ion bombardment of $500$ eV the deposited energy at the
depth of the interface (few eV) is insufficient to create Frenkel pairs (few tens of eV).
Hence this energy is dissipated into the lattice which leads to local
thermalization. On the other hand displaced atoms close to the surface of the film
if becomes not sputtered atoms, travel towards the interface as recoils.
These recoils create second or higher order generation of energetic atoms.
However, at this low incident ion energy regime the mean free path of these
atoms do not exceed the distance of few times of the lattice constant.
Hence their direct effect at the interface is negligible 
and these ballistic collisions of the recoils with the lattice atoms
can not cause TED in Pt/Ti.
The simulated nonlinear time evolution of $\langle R^2 \rangle$ indicates
the acceleration of particles. 
During the cascade intermixing of particles with the mass ratio of $\delta \approx 1$
no such nonlinear scaling of $\langle R^2 \rangle$ can be found.
This is because particles with nearly equal masses loose their kinetic energy
during elastic collisions and the lifetime of the cascade and spike period remains
short and which does not allow the evolution of $\langle R^2 \rangle \propto t^2$ time
scaling.
Hence from the conventional picture of
collisional cascades no intermixing acceleration can be expected.

  The projectile-to-target mass ratio could also play some role in the magnitude of
$F_D$ at the interface \cite{Nastasi}. Due to the large mass difference in the case of
$Ar^+$ $\rightarrow$ Pt (Pt/Ti) elastic collisions the energy loss is larger than
for $Ar^+$ $\rightarrow$ Ti impacts (Ti/Pt). Contrary to this we get 
the stronger IM in Pt/Ti.
It seems again that not simple binary collision effects the reason
of the asymmetry of IM.

\subsection{Thermal spike: heat of mixing}

 In recent publications we have already shown that we find the lack of the 
effect of heat of mixing $\Delta H$ of the corresponding alloy phases on ion-mixing in Ti/Pt \cite{Sule_NIMB04}.
This is in constrast with the predictions of the thermal spike model which
suggest that $\Delta H$ governs intermixing during the ion
bombardment of various bilayers \cite{Nastasi,AverbackRubia,TS}.
Repeating simulated ion-sputtering with varying $\Delta H$ in Pt/Ti, we find
again that the magnitude of ion-mixing is insensitive to the choice of
the $\Delta H$ which can be tuned by adjusting parameter $\xi$ (the preexponential
parameter in the attractive term in Eq. (1)) \cite{Sule_NIMB04}.
Even if $\Delta H \approx 0$ ($\xi =$0) strong IM occurs, although
in the alloy phase decomposition takes place.
Therefore the influence of thermodynamic driving forces can be ruled out and
we conclude that the thermal spike model might not be consistent with 
the occurrence of TED in Pt/Ti.

\subsection{Radiation enhanced diffusion: TED is athermal}

 Since neither the ballistic nor the TS model are consistent with our findings
in Pt/Ti we check whether other system parameters govern IM.
First, we discuss, whether the thermally activated radiation enhanced diffusion,
which ususally takes place after the TS period is responsible for the
asymmetry of IM in Pt/Ti and in Ti/Pt.
No temperature dependence has been found. The simulations provide nearly the same 
results in $\sim 0$ K and at room temperature events.
Large athermal experimental mixing rates ($k > 10^4$ $\hbox{\AA}^4$) have also been found for Ni/Al,
Cu/Al and Al/Mo bilayers by other groups \cite{Weber,Besenbacher,Ma}.
These results are inconsistent with the operation of RED induced TED in these materials \cite{Weber,Nastasi}.
Because of the insensitivity of TED in Pt/Ti to temperature effects we can also rule out
the influence of any conventional thermally activated vacancy and interstitial diffusion mechanisms \cite{Philibert}.

\subsection{The proposed mechanism}

 The radiation induced enhancement of IM in Ti/Pt has been studied in detail 
in ref. \cite{Sule_NIMB04}. The mass anisotropic interface stops
the light energetic particles which leads to overheating in the Ti phase in Ti/Pt.
This leads to the temporal decoupling of the IM of Pt and Ti
with the preferential IM of Pt \cite{Sule_NIMB04}.
The IM of Ti is delayed to the end of the TS period (retarded IM) \cite{Sule_NIMB04}.
In Ti/Pt the IM of Pt atoms take place, however, upwards to the film against the downward ion and recoil fluxes
which slows down IM.
In Pt/Ti, however, the preferential IM of Pt goes downwards to the substrate accelerated by the unidirectional incoming ions
and displaced energetic atoms.
Hence the heavier particle Pt behaves like a ballistic first diffuser in Pt/Ti and
as a slowed down particle by the counterflow of downward moving (from the film towards the substrate) energetic particles
of ion irradiation in Ti/Pt. 

  The ballistic preferential IM of Pt is governed by mass effect: the light energetic
particles (Ti) are backscattered at the heavy interface leading to the retardation of
IM of them \cite{Sule_NIMB04}.
%------------------------------------------------------
\begin{figure}[hbtp]
\begin{center}
\includegraphics*[height=4.cm,width=6.cm,angle=0.]{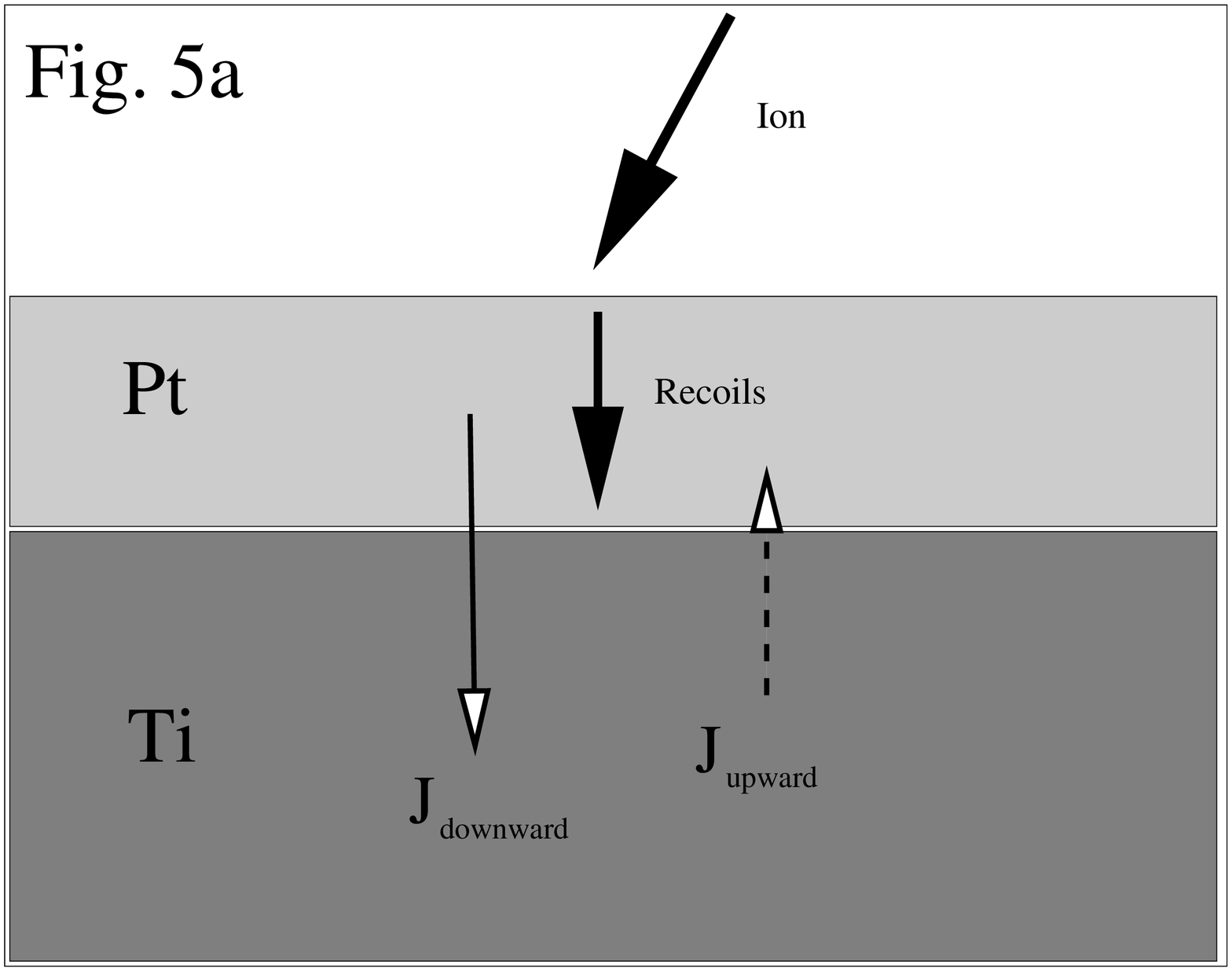}
\includegraphics*[height=4.cm,width=6.cm,angle=0.]{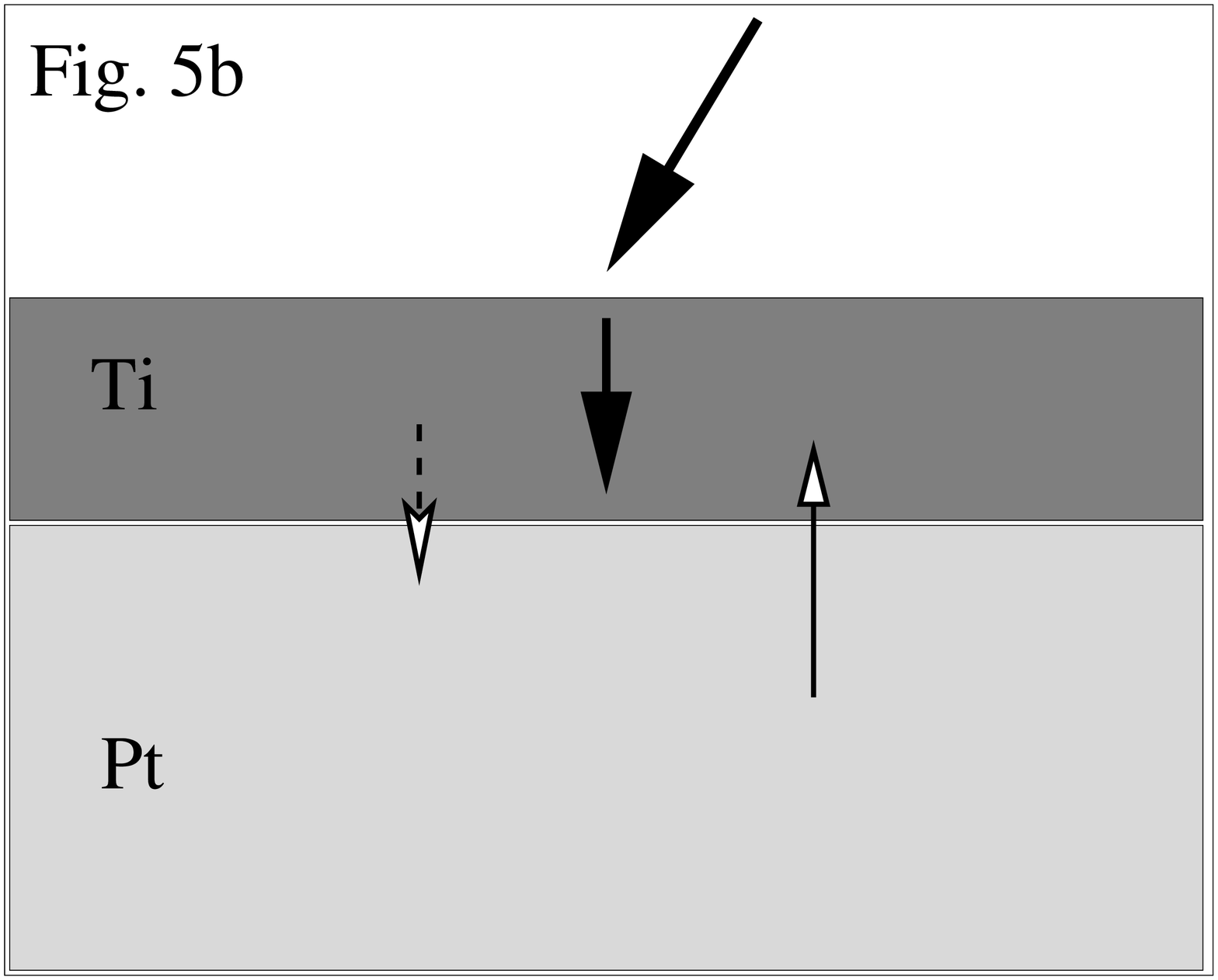}
\caption[]{
The schematic diagram of various ion-sputtering induced fluxes occur in
a general mass-anisotropic film/substrate bilayer.
Intermixing fluxes at the interface correspond to
$J_{\Uparrow,S}(\delta)$ (with upward arrow) and $J_{\Downarrow,F}(\delta)$ (downward arrow) while
the downward arrow in the film to the recoil flux of
$J_{\Downarrow}^{rec}$.
This configuration holds for $m_F \gg m_S$ where $m_F$ and $m_S$ denote the atomic masses of the film and
substrate constituents, respectively (FIG. 5a corresponds to  Pt/Ti).
Note the unidirectional fluxes of $J_{\Downarrow}^{rec}$ and $J_{\Downarrow,F}(\delta)$.
Moreover, fluxes $J_{\Uparrow,S}(\delta)$ and $J_{\Downarrow,F}(\delta)$ are
decoupled in time as it has been shown in ref. \cite{Sule_NIMB04}:
the mass current of Pt atoms $J_{\Downarrow,F}(\delta)$ goes predominantly
while $J_{\Uparrow,S}(\delta)$ is delayed by few ps.
 FIG. 5b (Ti/Pt): In this case fluxes $J_{\Downarrow}^{rec}$ and
$J_{\Uparrow,S}(\delta)$ (IM Pt atoms) are in opposite direction hence weakens each others effect
which leads to the suppression of IM.
This situation corresponds to the case when $m_F \ll m_S$.
}
\label{fluxes}
\end{center}
\end{figure}
%------------------------------------------------------
Also, the reversed flux of the light particles increases the energy density
in the Ti phase promoting the IM of the Pt atoms.

 Therefore we find the ion irradiation induced athermal preferential intermixing
of Pt atoms in Pt/Ti.
Two accelerating effects amplify each others effect: the mass effect induced preferential
interfacial mixing of energetic Pt atoms
and the downward fluxes of the incident ions and recoils contribute to ballistic IM in
Pt/Ti and to the emergence of nonlinear time scaling of $\langle R^2 \rangle$.
The mass anisotropy induced enhancement of preferential Pt interdiffusion occurs in both bilayers,
however, in Ti/Pt the fluxes of IM ballistic Pt atoms are somewhat slowed down by the
counterflow of the downward movement of recoils.
These simple reasonings explain the emergence of the asymmetry of IM with respect to
the interchange of film and substrate constituents.

\subsection{Phenomenology for $\delta$-driven TED}

 The phenomenological description of the asymmetric TED might help in understanding
and explaining the results obtained by experiment and MD simulations.
Following the Martin's theory of irradiation induced ballistic diffusion \cite{Martin}
and the Cahn-Hilliard theory of thermal diffusion \cite{Cahn}
the diffusion constant can be written formally as the sum of
thermally activated and ballistic (athermal) terms \cite{Martin}:
\be
D_{irrad}=D_{th}(T)+D_{\Downarrow}^{rec}.
\ee
The interdiffusion drift due to cascade mixing with recoils (hyperthermal 
particles) is given via
$D_{\Downarrow}^{rec}$. 
Within our picture of TED the mass effect induced amplification of IM
over the thermal and cascade mixing rates can be written as
\be
 D_{TED}(T,\delta)=D_{th}(T)+D_{\Downarrow}^{rec}+D_{enhan}(\delta),
\label{diff}
\ee
where $D_{th}(T)$ is a normal thermally activated diffusion constant,
where $T$ is the temperature in the thermal spike (irradiation induced molten phase) and
$D_{enhan}(\delta)$ is an enhancement term depending on $\delta$.
This model explains TED as an amplification of atomic intermixing
on top of radiation enhanced diffusion (thermally activated and collisional cascade ballistic interdiffusion
$D_{RED}=D_{th}(T)+D_{\Downarrow}^{rec}$.
Eq. ~\ref{diff} should give $D=D_{th}+D_{\Downarrow}^{rec}$ when $\delta \approx 1$.
It is not our intention in this paper to derive an explicit analytic expression which
could reproduce the MD results (nonlinear time scaling of $\langle R^2 \rangle$ in Pt/Ti)
as well as the experimental IM depth (long range diffusity tail) in the
concentration profile.
Simply we would like to explain in detail the mechanism of TED
within a phenomenological picture which helps understanding the amplification
of IM.

% According to the Einstein's relation \cite{Philibert,Michely}
%\be
%D_{TED}(\delta)=\frac{\langle R^2 \rangle(\delta)}{4 t},
%\ee
%the penetration depth of particles (the intermixing length or tail in the depth profile) 
%is given as follows,
%\be
% X = \sqrt{\langle R^2 \rangle} \approx \sqrt{4 t (D_{th} + D_{\Downarrow}^{rec} + D_{enhan}(\delta))}. 
%\ee

% To estimate the mass anisotropy induced increase of $X$ (broadening) one has to give an explicit formula for the mass enhancement diffusion term $D_{enhan}(\delta)$.

 {\em Unidirectional and counterflow atomic flux:} 
One possible way is to model mass effect by taking into account the counterflow
ballistic intermixing mass flows $J_{\Uparrow}$ and $J_{\Downarrow}$ normal to the surface appear during
ion-sputtering of bilayer systems.
As mentioned above particle flow of heavy particles (Pt) takes place downwards in Pt/Ti
and upwards in Ti/Pt.
%A counterflow of other not intermixing but downward moving energetic atoms also occur in Ti/Pt while a unidirectional
%mass current in Pt/Ti.
The total mass flow is the sum of these terms,
\be
 J(\delta)=J_{\Uparrow,A}(\delta)+J_{\Downarrow,B}(\delta)+J_{\Downarrow}^{rec},
\label{flux}
\ee
where $J_{\Uparrow,A}$ and $J_{\Downarrow,B}$ are the intermixing mass flow of constituents A and B through the
interface (the interface currents in a A/B bilayer, respectively.
The term $J_{\Downarrow}^{rec}$ is the downward flux of energetic particles occurs upon external
forced conditions (ion-sputtering) towards the interface. 
This term corresponds to the case 
when $\delta \approx 1$.
The $\delta$-induced downward flux amplification is $\Delta J(\delta)=J_{\Downarrow,B}(\delta)-J_{\Downarrow}^{rec}$,
which leads to superdiffusion.
The schematic view of the fluxes is shown in Fig.~\ref{fluxes} for a situation, when
the film component (F) is B and the substrate (S) is A.
Heavy particles have the tendency to IM preferentially over the light components \cite{Sule_NIMB04} which results in $J_{Pt} > J_{Ti}$.
In Pt/Ti, $J_{\Downarrow,B}(\delta)=J_{Pt}$, while in Ti/Pt
$J_{\Uparrow,A}(\delta)=J_{Pt}$.
It has also been shown recently that
the IM of the light and heavy components is decoupled in time by few ps due to the predominant
IM of the heavy atoms \cite{Sule_NIMB04}
which results in the robust amplification of the interface current of Pt
$J_{Pt} \gg J_{Ti}$.

 Intermixing atomic fluxes $J_{\Uparrow,A}$ and $J_{\Downarrow,B}$ (see Fig.~\ref{fluxes}) 
are created 
indirectly upon ion-sputtering, while flux $J_{\Downarrow}^{rec}$ appears directly upon
ion-bombardment.
However, $J_{\Uparrow,A}(\delta)$ and $J_{\Downarrow,B}(\delta)$ are directly tuned by $\delta$, while
flux $J_{\Downarrow}^{rec}$ is nearly independent of $\delta$.
This is because flux $J_{\Downarrow}^{rec}$ appears in the film while the IM fluxes
operate at the interface, where mass-anisotropy influences atomic transport directly.
This rationalizes the separation of ion-induced atomic fluxes into $\delta$-
dependent and independent terms.
If $\delta \rightarrow 1$, the sum of fluxes $J_{\Uparrow,A}+J_{\Downarrow,B} \approx 0$ vanishes 
and $J=J_{\Downarrow}^{rec}$. 
%Moreover, we can reorganize Eq. ~\ref{diff}.
The $\delta$-independent part of Eq. ~\ref{flux} is simply the
fluxes of cascade mixing term. 
Fluxes $J_{\Downarrow}^{rec}$ and $J_{\Downarrow,B}(\delta)$ appear nearly in the
same time (ballistic or cascade period), although flux $J_{\Downarrow,B}(\delta)$
is induced by $J_{\Downarrow}^{rec}$.
This is because the mass-anisotropic system gives an ultrafast response
to ion-irradiation and the $J_{\Downarrow,Pt}(\delta)$ flux 
in Pt/Ti and $J_{\Uparrow,Pt}(\delta)$ in Ti/Pt occur due to downwards recoils and
flux $J_{\Downarrow}^{rec}$.

 {\em $\delta$-driven particle acceleration:}
 $J_{\Downarrow}^{rec}$ and $J_{\Downarrow,B}(\delta)$ are unidirectional in Pt/Ti, hence the ion-irradiation induced flux $J_{\Downarrow}^{rec}$ accelerates Pt atoms in Pt/Ti and slows down $\delta$-driven IM in Ti/Pt because
$J_{\Downarrow}^{rec}$ and $J_{\Uparrow,A}$ are in contrary direction and
nearly counteract or weakens each others effect.
 The impinging ions always generate energetic particles with downward momentum which
is added to the momentum of $\delta$-driven heavy particles leading to the
huge amplification of atomic mobility of few intermixing atoms.
Hence we explain the huge interfacial broadening in Pt/Ti
by the cumulative effect of the two types of downward atomic mobilities.
%This leads to the concentration gradient 
%$\frac{\partial c}{\partial z}$.
%However, the counterflow of particles ($J_{\Uparrow,A}$) weakens the
%Hence the difference of upward and downward flows
%$\Delta J(\delta)$ is proportional to the mass effect.
%\be
% D_{ball}(\delta) \propto J_{\Uparrow}-J_{\Downarrow} \propto \delta.
%\ee
%When $\delta \rightarrow 1$, $\Delta J(\delta) \rightarrow 0$.

  The net intermixing mass flow can be described by the Fick's first law \cite{Philibert},
\be
 \Delta J(\delta) = -\frac{\partial c}{\partial z} (D_{\Uparrow}(\delta)-D_{\Downarrow}(\delta)),
\ee
where the concentration gradient occurs due to the mass anisotropy \cite{Sule_NIMB04}.
The thermal term does not lead to concentration difference and the broadening at the interface
is symmetrical, the IM of the components is nearly the same on both side of
the interface.
The athermal interdiffusion drift is the balance of the upward and downward
mass transport driven by $\delta$ and characterized
by the ballistic diffusion constants $D_{\Uparrow}(\delta)$ and $D_{\Downarrow}(\delta)$. 
%Because of the retarded IM and confinement of the light atoms the
%concentration gradient vanishes after the thermal spike \cite{Sule_NIMB04}.

%  We use in this study an expansion around $\delta_0=1$ for the
 Finally the amplification term of TED for the diffusion constant can be given as
\be
D_{enhan}(\delta)=-(J_{\Downarrow}(\delta)-J_{\Downarrow}^{rec}) \biggm(\frac{\partial c}{\partial z}\biggm)^{-1}.
%D_{\Downarrow}(\delta)=-{\Delta J_{\Downarrow}(\delta)}\biggm(\frac{\partial c}{\partial z}\biggm)^{-1} ~~~~~~~~~~~~~~~~~~~~ && \nonumber \\ =(D_{\Downarrow}^{rec} + D_{\Downarrow,B}) \biggm(\frac{\partial c}{\partial z}\biggm)^{-1} ~~~~~~~~~~~~~~~&& \nonumber \\ \approx D_{\Downarrow}^{rec}+a (\delta-\delta_0)+b ((\delta-\delta_0)^2+...
\ee
Therefore, if $\delta=1$, purely thermally activated diffusion (atomic transport)
takes place.
If $\delta \gg 1$, TED occurs (the amplification factor $D_{enhan}(\delta) \gg D_{RED}$).
When $\delta < 1$, TED features are suppressed, although
we get a stronger IM than in the case of mass isotropy $\delta \approx 1$
($D_{enhan}(\delta)  \approx D_{RED}$).

\subsection{Superdiffusion}

 Superdiffusive features, have never been reported before for intermixing, only
for e.g. random walk of adatoms and clusters (Levy flight) on solid surfaces 
\cite{anomalous,Levy,Luedtke,Michely}.
Transient mobility in the bulk has long been known only in collisional cascades
\cite{Nastasi,AverbackRubia,Gnaser}
during quantum diffusion of light particles \cite{Philibert}
or in shock loaded and stressed systems \cite{Philibert}.
However, these bulk phenomena are driven mostly by external stimulus (cascades, thermal spike, shock
loaded rearrangements) and can be characterized as driven systems \cite{Bellon,Martin}.
The only exception is the ultra low temperature ballistic diffusion of light particles
in the lattice which is driven by (intrinsic) quantum effects \cite{Philibert}. 
 Brockmann {\em et al.} has been attempted to interpret reacting particle
systems with front propagation driven by reaction-superdiffusion \cite{Brockmann}.

  The mass anisotropy driven TED in Pt/Ti can also be considered as a driven system.
 In $\delta > 1$ systems the strong mass anisotropy driven acceleration 
of the particles leads to superdiffusion of the heavier atoms.
However, the observed acceleration of the heavy particles is an intrinsic feature of
$\delta > 1$ systems.
The situation is somewhat similar to that found in the
 anomalous impurity diffusion of N in stainless steel \cite{Abrasonis} and
the observed large IM depths in various transition metal/Al diffusion
couples \cite{Buchanan} which could also be understood as super-interdiffusive processes.
In Ni/Al bi-, multi-, and marker layers an unsually high mixing rate has been
observed and which could not be understood within the picture of standard ion mixing
models \cite{Weber} as well as in our case in Pt/Ti.
These results suggest that anomalous and superdiffusive mass transport
could occur in various driven systems in which still unknown intrinsic system parameters
govern TED.
These parameters exist independently from the externally forced condition (ion bombardment).
The external perturbation of these systems is necessary, however, to induce
the transient atomic rearrangements.

 The superdiffusive features can be tuned by adjusting the mass ratio $\delta$ in Pt/Ti.
Setting in artificial mass isotropy, the nonlinear scaling of $\langle R^2 \rangle$ vanishes
(see Fig. ~\ref{R2}).
Hence mass anisotropy $\delta$ operates as a system parameter
($\delta = m_{film}/m_{substrate}$, where the atomic mass of the film has been divided
by the atomic mass of the substrate atoms),
If $\delta \gg 1$, super-interdiffusion appears, however, if $\delta \le 1$, 
IM slows down because of the counterflow of incident particles with IM Pt atoms.
In ref. \cite{Sule_PRB05} it has been shown, that the mixing efficiency $k/F_D$,
where $k$ is the mixing rate ($k=\langle R^2 \rangle/\Phi$) and $F_D$ and $\Phi$ are the deposited energy at the interface and the ion fluence, respectively \cite{AverbackRubia},
scales nonlinearily with $\delta$. 
At $\delta  < 0.33$ $k/F_D$ increases
abruptly. In that article we studied the ion mixing of A/B bilayers with $\delta  \le 1$. However,
we did not study the inverted systems (B/A,  $\delta  > 1$, film atoms are much heavier than the
substrate one).
In B/A systems in general, in which $\delta \gg 1$, mass anisotropy driven superdiffusion
might occur on the basis of the present results.
This can be shown by computer experiments:
if we simulate ion-mixing in mass isotropic systems, such as Co/Cu e.g., we get
a very weak interfacial mixing. However, if we set in artificially $\delta \gg 1$, 
strong IM takes place as in Pt/Ti.
We reach the conclusion that the simple system parameter $\delta$ governs
the enhancement of IM in mass anisotropic bilayers.
%Moreover, if the much heavier constituent is in the film, than
%the transient injection of these heavy particles take place
%into the substrate via superdiffusion (that is the ballistic flight of few energetic atoms).
In the inverted case (A/B), when the lighter constituents are placed in the film,
no transient enhanced intermixing occurs.

\section{Conclusions}

 The most important findings are the following:

  (i) {\em Mass effect and asymmetric mixing:} We find a robust mass effect on interfacial mixing in Pt/Ti which
supports our finding published in ref. \cite{Sule_PRB05} in which a
strong mass effect on ion-beam intermixing (IM) has been found for various mass-anisotropic bilayers.
%However, the magnitude of mass anisotropy induced intermixing strongly depends on the succession of the film and
%substrate in the Ti/Pt bilayer.
%Using molecular dynamics simulations in Pt/Ti a strong IM, while in Ti/Pt a much weaker one has been found
%in accordance with our depth profiling measurements \cite{Sule_JAP07}.
In order to increase the credibility of the employed computational approach,
we fitted the crossinteraction atomic interaction potential to that of obtained from first principles
calculations.

 (ii) {\em Nonlinear time scaling:} We find that the sum of the squares of atomic displacements through the anisotropic interface
($\langle R^2 \rangle$)
scales nonlinearly in Pt/Ti ($\langle R^2 \rangle \propto t^2$) as a function of the time (and the ion-number fluence)
as shown in Fig. ~\ref{R2}.
The nonlinear time scaling of $\langle R^2 \rangle$ together with the
long range (high diffusity) tail in the AES profile shown in ref. \cite{Sule_JAP07} might support the operation of a superdiffusive transport
(athermal) process of Pt atoms in Pt/Ti.
In Ti/Pt a nearly linear scaling ($\langle R^2 \rangle \propto t$) is found.
The lack of
a tail in the AES concentration profile for Ti/Pt (\cite{Sule_JAP07}) 
is explained by the suppression of the preferentail IM of Pt into
the Ti phase due to the counterflow fluxes of downward moving recoils and
the upwards mobility Pt atoms.

 (iii) {\em Preferential mixing of Pt and further acceleration:} The atomistic mechanism
 of the asymmetry and TED in Pt/Ti is the following:
The mass-anisotropy induced predominant intermixing 
of the heavier Pt atoms into the Ti phase has been found in both materials
in accordance with previous findings
\cite{Sule_NIMB04}.
The new finding is that
in Ti/Pt the Pt atoms slow down during IM because of
the slowing down effect of the counterflow of the incident energetic particles (recoils) during the cascade
period.
In Pt/Ti we find the contrary situation: the IM Pt atoms are accelerated by the
unidirectional current of the hyperthermal particles (downward Pt recoils), hence
 the originally already
preferentially intermixing Pt atoms even further accelerated.
Hence the superdiffusion of these particles is driven by double acceleration:
first driven by the mass ratio induced preferential transport and additionally by the
ballistic particles in the collisional cascade with downward mobility
(their momentum directed from the film towards the substrate). 
Such kind of an acceleration of particles in the bulk has never been reported before
at best of our knowledge,
although, this situation can be rather general in
$\delta > 1$ systems.

 The phenomenon is governed by the intrinsic system parameter mass anisotropy and not
or only weakly influenced by external parameters, such as the ion specie, projectile to target
mass ratio, the ion energy above few hundred eV or the external temperature.
This is because the process requires only a unidirectional current of
energetic particles with different origin, which is present in the system nearly independently from the
parameters mentioned above.
In this sense this unique feature of the superdiffusive TED makes it
different from other ballistic atomic transport processes occuring in e.g.
collisional cascades which strongly influenced by external parameters.

 (iv) {\em Unconventional mechanism:} We conclude that the observed and simulated long range depth distribution of Pt atoms
in the Ti phase of Pt/Ti cannot be understood by any established mechanisms of radiation-enhanced
diffusion (RED).
We find that the occurrence of the long range diffusity tail, whose penetration depth
exceeds the ion range could be understood as a superdiffusive process in the bulk.
 Moreover, normally, RED could not lead to the asymmetry of IM. That is because intermixing
in the collisional cascade is normally insensitive to the succession of the layers.
Also, the thermal spike (which is rather short at this ion energy) in principle
could not provide asymmetric IM.
 Instead we speculate on the possible operation of accelerative effects which could enhance atomic mixing and penetration.
The divergence of $\langle R^2 \rangle$ clearly indicates that accelerative effects are present
in the lattice leading to ballistic transport.
$\langle R^2 \rangle \propto t^2$ is the signature of ballistic atomic transport \cite{anomalous}.
The time evolution of nonlinear time scaling of $\langle R^2 \rangle$ requires the sufficiently
large number of ballistic particles with large mean free path during most of the cascade events.
During the "normal" cascade events of Ti/Pt 
$\langle R^2 \rangle$ does not exceed linear scaling.
Hence a specific mechanism might come into play which speeds up Pt particles
in the Ti bulk or at the interface during the collisional cascade period in Pt/Ti.

 (v) {\em Unidirectional energetic atomic fluxes:} The $\delta$-driven TED has also been explained using a phenomenological model
based on a simple Fickian model.
This helps in explaining the occurrence of the established accelerative effect on IM
driven by unidirectional atomic fluxes.
The fluxes of energetic particles can be separated into $\delta$-dependent
and independent terms. The corresponding diffusion constant is
constructed as a sum of thermally activated, cascade (recoil) mixing and $\delta$-dependent
parts. The unidirectional fluxes in Pt/Ti leads to huge amplification 
of downwards fluxes while in Ti/Pt these fluxes are in counterflow direction
leading to the weakening of the upwards intermixing fluxes of Pt atoms.

 (vi) {\em The $\delta$-driven AAT might be a general mechanism:} 
Finally we conclude that the established mechanism of TED might not be
a specific one in nature.
{\em Nanoscale mass-anisotropy} induced AAT could be a general feature of various multicomponent systems
and could occur during the ion-beam processing (ion-sputtering, dopant implantation or
ion-beam deposition)
in various thin film multilayers, nanoinclusions, nanoislands, quantumdots embedded in
light atomic host matrices (substrates) or in quantum well structures.
Such kind of nanostructures and processing technologies are widely used in the production of
heterostructure nanodevices \cite{Schukin} or
magnetic nano-objects which are of high current interest due to 
numerous potential applications in various fields \cite{nano}. 

  In particular,
TED has been found in nonstochiometric AlAs/GaAs quantum well
structures \cite{GaAs} or in AsSb/GaSb superlattices \cite{AsSb}
which can be considered as nanoscale mass-anisotropic systems with
well-defined interfaces.
The reported AAT in these systems could also be, at least partly, due to
$\delta$-driven AAT mechanism.
Also, possibly there are couple of other systems in which 
the $\delta$-driven AAT could take place. Just to mention
few examples: sputter deposition of transition metals on Al \cite{Buchanan,ptonal},
low energy cluster deposition and pinning on various substrates \cite{ptonal}.
In these system it has already been shown by computer simulations that a $\delta$-driven AAT
mechanism plays a significant role during thin film growth \cite{ptonal}.

   The most important structural condition which must be fullfield is that  
the occurrence of $\delta$-driven TED requires the presence of a mass-anisotropic interface
with $\delta \gg 1$.
Hence when the primary goal is the production of nanostructured surfaces
sharp interfaces are required for the efficient operation of nanodevices.
According to our results,
however, nanoscale mass-anisotropy might deteriorate the sharpness
of mass-anisotropic interfaces, especially when ion-sputtering have been
used during the processing of the nanostructured surfaces
and interfaces.
One possible way of avoiding $\delta$-driven interface broadening is
the construction of $\delta \le 1$ multilayer thin films which
do not allow the amplification of intermixing and even
lead to the suppression of interdiffusion as it has been found
in the Ti/Pt bilayer.
The better understanding of $\delta$-driven AAT could help in the
more efficient production of nanothin films with sharp interfaces.

{\scriptsize
This work is supported by the OTKA grant F037710
from the Hungarian Academy of Sciences.
We wish to thank to K. Nordlund 
for helpful discussions and constant help.
The work has been performed partly under the project
HPC-EUROPA (RII3-CT-2003-506079) with the support of
the European Community using the supercomputing 
facility at CINECA in Bologna.
The help of the NKFP project of 
3A/071/2004 is also acknowledged.
The G03 code is available at NIIF center at Budapest. 
}

\end{document}